\documentclass[Times,4pt,aps,pra,twocolumn,showpacs,amsmath,amssymb,floatfix,footinbib,superscriptaddress]{revtex4-2}
\usepackage{amsmath,natbib,graphicx,subfigure,amssymb,graphics,amsmath,mathrsfs,CJK,color,comment}
\usepackage{multirow,fancyhdr,color,bm,tabularx,psfrag,geometry,dcolumn,datetime,physics,booktabs}
\usepackage[colorlinks=true,linkcolor=blue,urlcolor=blue,citecolor=blue]{hyperref}
\usepackage[mathlines]{lineno}
\def\footnoterule{\kern -1mm \hrule width 5.8cm \kern 2.2mm}
\usepackage{tikz,xcolor,hyperref}
\definecolor{lime}{HTML}{A6CE39}
\DeclareRobustCommand{\orcidicon}{%
    \begin{tikzpicture}
    \draw[lime, fill=lime] (0,0)
    circle [radius=0.16]
    node[white] {{\fontfamily{qag}\selectfont \tiny ID}};\draw[white, fill=white] (-0.0625,0.095)
    circle [radius=0.007];
    \end{tikzpicture}
    \hspace{-2mm}}
\foreach \x in {A, ..., Z}
{\expandafter\xdef\csname orcid\x\endcsname{\noexpand\href{https://orcid.org/\csname orcidauthor\x\endcsname}{\noexpand\orcidicon}}}

\begin{document}


\title{Different roles of quantum interference in a quantum dot photocell with two intermediate bands}

\author{Shun-Cai Zhao\orcidA{}}
\email[Corresponding author: ]{zhaosc@kmust.edu.cn }
\affiliation{Department of Physics, Faculty of Science, Kunming University of Science and Technology, Kunming, 650500, PR China}
\author{Jing-Yi Chen}
\affiliation{Department of Physics, Faculty of Science, Kunming University of Science and Technology, Kunming, 650500, PR China}
\author{Xin Li}
\affiliation{Department of Physics, Faculty of Science, Kunming University of Science and Technology, Kunming, 650500, PR China}


\begin{abstract}
It is generally believed that quantum interference can improve the transport of photo-generated carriers in a photocell, thereby improve the photoelectric conversion efficiency. In this work, we explicitly explore different roles of quantum interferences in the photoelectric conversion efficiency in a quantum dot (QD) photocell with two intermediate bands. The increasing transition rates from different charge transport channels bring out first increasing, then decreasing, and then monotonically decreasing photoelectric conversion efficiencies. And the photoelectric conversions increase with quantum coherence generated by the upper transition rates owing to their robust quantum interference. However, the conversion efficiency decrease with the quantum interference induced by two lower-transition rates due to the shortened population lifetime in the intermediate bands. These results provide insight into different roles of quantum interferences in photoelectric conversion efficiency, and may provide some artificial strategies to achieve efficient photoelectric conversion via the adjusted quantum interferences in a QD photocell with multi-intermediate bands.
\begin{description}
\item[PACS numbers]42.50.Gy
\item[Keywords] Photoelectric conversion efficiency; quantum dot photocell; lower and upper interband transition rates;
\end{description}
\end{abstract}
\maketitle
\section{INTRODUCTION}
~~~~The efficiency of photoelectric conversion in a single bandgap solar cell is fundamentally limited to 31\(\%\) because of the efficiency loss caused by the detailed balance limit\cite{Shockley1961}. And the types of efficiency loss can be divided into two distinct categories under one sun illumination. One is the extrinsic losses, such as series resistance\cite{Handy1967Theoretical,SAGLAM1996Series,Kar2015Series}, parasitic recombination and contact shadowing\cite{Tobler2009A,Kajiyama2012Recent}, they are theoretically avoidable and consequently are not considered in fundamental limiting efficiency. The other is the intrinsic losses, such as non-absorption of photons with energy below the bandgap\cite{Luque2013Interband,Tian2015Non,Kum2018Two}, thermalisation loss\cite{Richards2006Enhancing,CONIBEER2010Modelling,Hirst2011Hot,Conibeer2012Hot} due to strong interaction between excited carriers and lattice phonons, and emission loss according to Kirchoff`s law\cite{Graziani2004The,Yi2018Maximal}, they are unavoidable in device design and will still be present in an idealised solar cell\cite{Henry1980Limiting}. Therefore, the intermediate band solar cell \cite{Klimov2004High,doi:10.1063/1.3600702,Luque2012Understanding} were proposed to overcome some fundamental limitation by introducing a radiative efficiency but electrically isolated band between the conduction and valence bands\cite{doi:10.1063/1.4941793,Beattie2017Quantum,Semonin2011Peak}.

Recently, some researches manifest that\cite{Scully2010Quantum,Svidzinsky2011Enhancing,Zhao2019} the detailed balance can be broken by quantum coherence induced by the intermediate band transitions, which achieves the enhanced quantum efficiency in a photocell\cite{Daryani2017High}.
As far as we know, the inter-band transition rates linked to quantum interference have been confirmed to be sensitive to electron-hole pair (exciton) interaction and quantum confinement\cite{PhysRevB.38.9797}. Hence, the electron-hole correlation was used to tune inter-band and intra-band optical transition rates in semiconductor nanostructures, and some experiments\cite{Ithurria2011Colloidal,Achtstein2012Electronic,Tighineanu2016Single} have revealed band edge transition rates reaching 10 ns\(^{-1}\)- about two orders of magnitude faster than those in strongly confined quantum dots. Moreover, the photon ratchet scheme was also introduced to increase the lifetime of charge carriers in the intermediate state\cite{Yoshida2012Photon} in an intermediate band solar cell for efficient photovoltaic energy conversion.

However, in a multi-band QD photovoltaic cell, such as a single-band gap QD photovoltaic cell incorporated two intermediate bands, two different quantum interferences will be generated from two different charge transfer channels. Are the roles of different quantum interferences the same in the photoelectric conversion efficiency? Does the different quantum interferences always promote the photoelectric conversion efficiency? Thereupon, revealing the underlying physical mechanisms of different quantum interferences is beneficial to answer the above questions and beneficial to precisely control the quantum interference for artificial high-efficiency multi-level QD photovoltaic cells.

\begin{widetext}
\begin{eqnarray}
&&\left(\Delta_p+\delta-k_pv+i\frac{\Gamma_{eg}}{2}\right)\rho^{(1)}_{eg}+\frac{\Omega_{c}}{2}\rho^{(1)}_{rg}-\frac{\Omega_{p1}}{2}\left(2\rho^{(1)}_{ee}+\rho^{(1)}_{rr}\right)-\frac{\Omega_{p2}}{2}\left(2\rho^{(0)}_{ee}+\rho^{(0)}_{rr}-1\right)=0 \\
&&\left(\Delta_2+\delta-\Delta kv+i\frac{\Gamma_{2}}{2}\right)\rho^{(1)}_{rg}+\frac{\Omega_{c}}{2}\rho^{(1)}_{eg}-\frac{\Omega_{p1}}{2}\rho^{(1)}_{re}
-\frac{\Omega_{p2}}{2}\rho^{(0)}_{re}=0 \\
&&\left(\Delta_c+\delta+k_cv+i\frac{\Gamma_3}{2}\right)\rho^{(1)}_{re}-\frac{\Omega_{p1}}{2}\rho^{(1)}_{rg}-\frac{\Omega_{c}}{2}(\rho^{(1)}_{rr}-\rho^{(1)}_{ee})=0 \\
&&\left(\Delta_p-\delta-k_pv-i\frac{\Gamma_{eg}}{2}\right)\rho^{(1)}_{ge}+\frac{\Omega_{c}}{2}\rho^{(1)}_{gr}-\frac{\Omega_{p1}}{2}(2\rho^{(1)}_{ee}+\rho^{(1)}_{rr})=0 \\
&&\left(\Delta_{2}-\delta-\Delta kv-i\frac{\Gamma_2}{2}\right)\rho^{(1)}_{gr}+\frac{\Omega_{c}}{2}\rho^{(1)}_{ge}-\frac{\Omega_{p1}}{2}\rho^{(1)}_{er}=0\\
&&\left(\Delta_{c}-\delta+k_cv-i\frac{\Gamma_3}{2}\right)\rho^{(1)}_{er}+\frac{\Omega_{c}}{2}(\rho^{(1)}_{ee}-\rho^{(1)}_{rr})-\frac{\Omega_{p2}}{2}\rho^{(0)}_{gr}-\frac{\Omega_{p1}}{2}\rho^{(1)}_{gr}=0 \\
&&\frac{\Omega_{c}}{2}\left(\rho^{(1)}_{re}-\rho^{(1)}_{er}\right)-\delta\rho^{(1)}_{rr}-i\Gamma_{2}\rho^{(1)}_{rr}=0 \\
&&\frac{\Omega_{p1}}{2}\left(\rho^{(1)}_{eg}-\rho^{(1)}_{ge}\right)+\frac{\Omega_{c}}{2}\left(\rho^{(1)}_{er}-\rho^{(1)}_{re}\right)+\frac{\Omega_{p2}}{2}\rho^{(0)}_{ge}-\delta\rho^{(1)}_{ee}-i\Gamma_{eg}\rho^{(1)}_{ee}+i\Gamma_{re}\rho^{(1)}_{rr}=0 \\
\nonumber
\end{eqnarray}
\end{widetext}

Therefore, this work investigates quantum coherence dependent photoelectric conversion in a QD photocell with two intermediated bands. It shows, that the quantum coherence generated from the upper and lower intermediate band transitions plays opposite roles in the photoelectric conversion efficiency due to their different underlying physical mechanisms. These results reveal the exact identity of quantum coherence and will help to enhance the photoelectric conversion efficiency in artificial QD photovoltaic cells.

\section{The master equations description of ideal photocell model}

\begin{figure}[htp]
\centerline{\includegraphics[width=0.25\columnwidth]{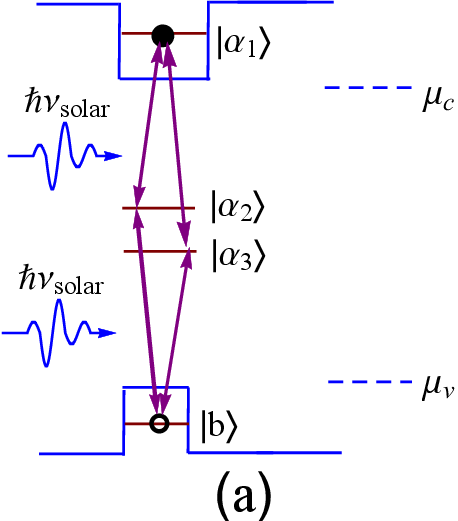}~~~~~~~~~~~~~\includegraphics[width=0.3\columnwidth]{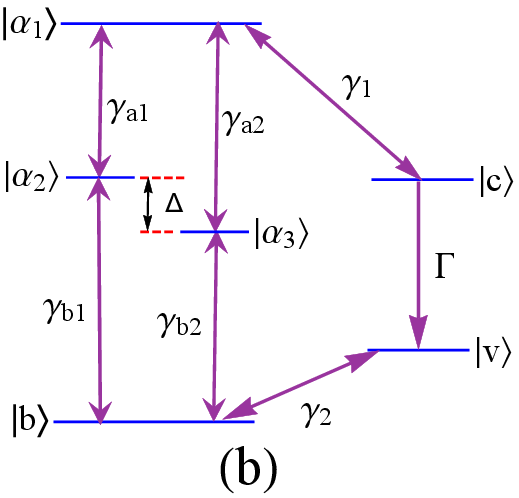}}
\caption{(Color online) Schematic of a QD solar cell. (a) Single-gap QD cell with two intermediate states (\(|\alpha_{2}\rangle\),\(|\alpha_{3}\rangle\)), two dashed lines indicate chemical potentials of the valence and conduction bands \(\mu_{c}\), \(\mu_{v}\), respectively. Solar radiation drives two upper interband transitions \(|\alpha_{1}\rangle\)\(\leftrightarrow\)\(|\alpha_{2,3}\rangle\), and two lower interband transitions \(|\alpha_{2,3}\rangle\)\(\leftrightarrow\)\(|b\rangle\). (b) Corresponding energy-level diagram for the QD solar cell. The absorbed photons drive transitions through two charge transfer channels \(|\alpha_{1}\rangle\) \(\leftrightarrow\) \(|\alpha_{2}\rangle\) \(\leftrightarrow\) \(|b\rangle\), \(|\alpha_{1}\rangle\) \(\leftrightarrow\) \(|\alpha_{3}\rangle\) \(\leftrightarrow\) \(|b\rangle\). Transitions \(|\alpha_{1}\rangle\)\(\leftrightarrow\)\(|c\rangle\) and \(|v\rangle\)\(\leftrightarrow\)\(|b\rangle\) are driven by the ambient thermal phonons. A load is connected to the two levels \(|c\rangle\) and \(|v\rangle\).}
\label{fig1}
\end{figure}

This proposed model of QD photovoltaic cell consists of an array of quantum dots and has two intermediate bands \(|\alpha_{2}\rangle\), \(|\alpha_{3}\rangle\), the energy gap between the valence and the conduction bands is divided into several sub-bands (Seen in Fig.\ref{fig1} (a)). Hence, more photons with lower energy can be absorbed and the excited electrons are transferred from the valence band to conduction band via the intermediate levels, then contributing to the useful work and enhancing photocurrent. These processes are efficient if the intermediate bands have more electronic population effectively transferred to the conduction band. For the sake of simplicity, photons with energy above the band-gap between the conduction and valence bands will not be considered in this QD photovoltaic cell model.

We assume that solar photons are resonantly absorbed by the upper interband transition with energy-gaps \(E_{\alpha_{1}}\)-\(E_{\alpha_{2,3}}\) or by the lower interband transitions \(|\alpha_{2,3}\rangle\)\(\leftrightarrow\)\(|b\rangle\) with energy-gaps \(E_{\alpha_{2,3}}\)-\(E_{b}\), respectively. When one photon with energy \(E_{\alpha_{2}}\)-\(E_{b}\) is absorbed, and the electron is excited from the valence band level \(|b\rangle\) to impurity level \(|\alpha_{2}\rangle\). Then another photon with energy \(E_{\alpha_{2}}\)-\(E_{\alpha_{1}}\) promotes the electron from level \(|\alpha_{2}\rangle\) to the conduction band level \(|\alpha_{1}\rangle\). The same two-photon excitation process can also take place from another transition channel \(|b\rangle\) \(\rightarrow\) \(|\alpha_{3}\rangle\) \(\rightarrow\) \(|\alpha_{1}\rangle\), simultaneously (Seen in Fig.\ref{fig1} (a)). These multiple absorptive pathways and re-emission will result in different quantum coherence in the proposed QD photocell. Levels  \(|c\rangle\) and \(|v\rangle\) corresponding to the conduction and valence bands are connected to a load modeled by the decay rate \(\Gamma\). \(\gamma_{ai}\), \(\gamma_{bi}\) (i=1,2) are the corresponding spontaneous decay rates of carriers between these transitions, \(|\alpha_{1}\rangle\) \(\leftrightarrow\) \(|\alpha_{2, 3}\rangle\) and \(|\alpha_{2, 3}\rangle\) \(\leftrightarrow\) \(|b\rangle\), and \(\gamma_{1, 2}\) are the decay rates of transitions  \(|\alpha_{1}\rangle\) \(\leftrightarrow\) \(|c\rangle\) and \(|v\rangle\) \(\leftrightarrow\) \(|b\rangle\), respectively (Seen in Fig.\ref{fig1} (b)). \(\Delta=\omega_{1}-\omega_{2}\) is the gap of the levels \(|\alpha_{2}\rangle\) and \(|\alpha_{3}\rangle\). And this system was assumed to interact with thermal phonon reservoir via the transitions  \(|\alpha_{1}\rangle\) \(\leftrightarrow\) \(|c\rangle\) and \(|v\rangle\) \(\leftrightarrow\) \(|b\rangle\). The direct radiative transition between \(|\alpha_{1}\rangle\) and \(|b\rangle\) will not be considered in this proposed QD model owing to its relatively small value caused by two incorporated intermediate bands in the real experimental research and being discussed in previous work\cite{Konstantin2013Increasing}. In the interaction picture, Hamiltonian for this system in rotating wave approximation is read as follows,

\begin{eqnarray}
&\hat{V}(t)=&\hbar \{\sum_{j} g_{j}e^{i(\omega_{12}-\nu_{j})t}\hat{\sigma}_{1} \hat{a}^{\dag}_{j}+\sum_{k}g_{k}e^{i(\omega_{13} -\nu_{k})t}\hat{\sigma}_{2} \hat{a}^{\dag}_{k}+\nonumber\\
       && \sum_{m} g_{m} e^{i(\omega_{2b}-\nu_{m})t}\hat{\sigma}_{3} \hat{a}^{\dag}_{m}+ \sum_{n} g_{n} e^{i(\omega_{3b}-\nu_{n})t}\hat{\sigma}_{4} \hat{a}^{\dag}_{n}+ \nonumber\\
       && \sum_{o}g_{o}e^{i(\omega_{1c}-\nu_{o})t}\hat{\sigma}_{5}\hat{b}^{\dag}_{o} +\sum_{p}g_{p} e^{i(\omega_{bv}-\nu_{p})t}\hat{\sigma}_{6} \hat{b}^{\dag}_{p}\}+h.c.\\\nonumber
\label{eq.1}
\end{eqnarray}
\noindent  where \(g_{j,k,m,n,o,p}\) are the coupling constants for the corresponding transitions, respectively. \(\hbar\omega_{ij}=E_{i}-E_{j}\) indicates the energy spacing between levels \(|i\rangle\) and \(|j\rangle\), \(\nu_{i}\) is the photon or phonon frequency. \(\hat{\sigma}_{i}(i=1,2,3,4,5,6)\) is the Pauli lowering operator, and \(\hat{a}^{\dag}_{j,k,m,n}\), \(\hat{b}^{\dag}_{o, p}\) are the photon and phonon creation operators, respectively. The equation of motion for the electronic density operator \(\hat{\rho}\) is deduced in the interaction picture as following,
\begin{equation}
\dot{\hat{\rho}}(t)=-\frac{1}{\hbar^{2}}Tr_{R} \int^{t}_{t_{0}}[\hat{V}(t),[\hat{V}(t'),\hat{\rho}(t')\bigotimes\hat{\rho}_{R}(t_{0})]]dt',
\label{eq.2}
\end{equation}
\noindent where the density operator \(\hat{\rho}_{R}\) describes the phonon thermal reservoirs. Invoking the conventional second-order perturbative treatment with respect to \(\hat{V}(t)\), the Born-Markov approximation, and the Weisskopf-Wigner approximation, Eq.(\ref{eq.2}) can be described by two divided Lindblad-type super-operators:
\begin{eqnarray}
&\mathfrak{R}_{ij(k)}\{\rho\}\!=\!&\sum_{k=1,2}\sum_{i=a,b}\sum^{2}_{j=1}\frac{\gamma_{ij(k)}}{2}[(n_{ij(k)}+1)(2\hat{\sigma}_{ij(k)}\hat{\rho}\hat{\sigma}^{\dag}_{ij(k)} \nonumber\\
&&-\hat{\sigma}^{\dag}_{ij(k)}\hat{\sigma}_{ij(k)}\hat{\rho}-\hat{\rho}\hat{\sigma}^{\dag}_{ij(k)}\hat{\sigma}_{ij(k)})+n_{ij(k)}(2\hat{\sigma}^{\dag}_{ij(k)}\hat{\rho}\hat{\sigma}_{ij(k)}\nonumber\\
&&-\hat{\sigma}_{ij(k)}\hat{\sigma}^{\dag}_{ij(k)}\hat{\rho}-\hat{\rho}\hat{\sigma}_{ij(k)}\hat{\sigma}^{\dag}_{ij(k)})],\\ &\mathfrak{R}_{\Gamma}\{\rho\}\!=\!&\Gamma[(2\hat{\sigma}_{\beta\alpha}\hat{\rho}\hat{\sigma}^{\dag}_{\beta\alpha}-\hat{\sigma}^{\dag}_{\beta\alpha}\hat{\sigma}_{\beta\alpha}\hat{\rho}-\hat{\rho}\hat{\sigma}^{\dag}_{\beta\alpha}\hat{\sigma}_{\beta\alpha})].
\end{eqnarray}
Hence, the reduced matrix density operators in the master equations describing the electronic system interacting with radiation and phonon reservoirs are obtained in the following,
\begin{eqnarray}
&\dot{\rho_{11}}\!=\!&-\gamma_{a1}[(1+n_{a1})\rho_{11}-n_{a1}\rho_{22}]-\gamma_{a2}[(1+n_{a2})\rho_{11}-n_{a2}\rho_{33}]\nonumber\\
                   &&-P_{1}\sqrt{\gamma_{a1}\gamma_{a2}}(n_{a1}+n_{a2})Re[\rho_{23}]-\gamma_{1}[(1+n_{2})\rho_{11}-n_{2}\rho_{cc}],\nonumber\\
&\dot{\rho_{22}}\!=\!&-\gamma_{b1}[(1+n_{b1})\rho_{22}-n_{b1}\rho_{bb}]+\gamma_{a1}[(1+n_{a1})\rho_{11}-n_{a1}\rho_{22}]\nonumber\\
                   &&-P_{1}\sqrt{\gamma_{a1}\gamma_{a2}}n_{a2}Re[\rho_{23}]-P_{2}\sqrt{\gamma_{b1}\gamma_{b2}} (1+n_{b2})Re[\rho_{23}], \label{eq.3}\\
&\dot{\rho_{33}}\!=\!&-\gamma_{b2}[(1+n_{b2})\rho_{33}-n_{b2}\rho_{bb}]+\gamma_{a2}[(1+n_{a2})\rho_{11}-n_{a2}\rho_{33}]\nonumber\\
                   && -P_{1}\sqrt{\gamma_{a1}\gamma_{a2}}n_{a1}Re[\rho_{23}]-P_{2}\sqrt{\gamma_{b1}\gamma_{b2}} (1+n_{b1})Re[\rho_{23}],\nonumber\\
&\dot{\rho_{23}}\!=\!&-i\Delta\rho_{23}-\frac{1}{2}\rho_{23}[\gamma_{b1}(1+n_{b1})+\gamma_{b2}(1+n_{b2})+\gamma_{a1}n_{a1}+\gamma_{a2}n_{a2}]\nonumber\\
                  &&-\frac{1}{2}P_{1}\sqrt{\gamma_{a1}\gamma_{a2}}[n_{a1}\rho_{22}+n_{a2}\rho_{33}-(n_{a1}+n_{a2}+2)\rho_{11}]\nonumber\\
                  &&-\frac{1}{2}P_{2}\sqrt{\gamma_{b1}\gamma_{b2}}[(1+n_{b1})\rho_{22}+(1+n_{b2})\rho_{33}-(n_{b1}+n_{b2})\rho_{bb}] \nonumber\\
&\dot{\rho_{cc}}\!=\!&\gamma_{1}[(1+n_{1})\rho_{11}-n_{1}\rho_{cc}]-\Gamma \rho_{cc},\nonumber\\
&\dot{\rho_{vv}}\!=\!&\Gamma\rho_{cc}-\gamma_{2}(1+n_{2})\rho_{vv}+\gamma_{2}n_{2}\rho_{bb}. \nonumber
\end{eqnarray}

\noindent where \(n_{a1,2}\) and \(n_{b1,2}\) are the mean number of solar photons with energy (\(E_{\alpha 1}-E_{\alpha 2,3}\)) and (\(E_{\alpha 2,3}-E_{b}\))  between the transition \(|\alpha_{1}\rangle\) \(\leftrightarrow\) \(|\alpha_{2, 3}\rangle\) and \(|\alpha_{2, 3}\rangle\) \(\leftrightarrow\) \(|b\rangle\),

\begin{eqnarray}
&n_{a1}\!=\!\frac{1}{exp(\frac{E_{\alpha 1}-E_{\alpha 2}}{K_{B}T_{s}})-1},~~n_{a2}\!=\!\frac{1}{exp(\frac{E_{\alpha 1}-E_{\alpha 3}}{K_{B}T_{s}})-1},\nonumber\\
&n_{b1}\!=\!\frac{1}{exp(\frac{E_{\alpha 2}-E_{b}}{K_{B}T_{s}})-1},~~n_{b2}\!=\!\frac{1}{exp(\frac{E_{\alpha 3}-E_{b}}{K_{B}T_{s}})-1}, ~~\label{eq.4}
\end{eqnarray}

\noindent They represent the robustness of the solar radiations with the solar temperature \(T_{s}\). And \(n_{1,2}\) are the mean number of thermal phonons with energy (\(E_{\alpha 1}-E_{c}\)) and (\(E_{v}-E_{b}\)),

\begin{eqnarray}
n_{1}=\frac{1}{exp(\frac{E_{\alpha 1}-E_{c}}{K_{B}T_{a}})-1},~~~~~n_{2}=\frac{1}{exp(\frac{E_{v}-E_{b}}{K_{B}T_{a}})-1}.
\label{eq.5}
\end{eqnarray}

\noindent respectively. \(T_{a}\) is ambient circumstance temperature. And \(P_{1}\) and \(P_{2}\) in Eq.(3) are the dipole alignment factors of their corresponding dipole matrix elements with the following expressions,

\begin{eqnarray}
P_{1}=\frac{\vec{p}_{a1}\cdot \vec{p}_{a2}}{ \vert \vec{p}_{a1} \vert\cdot\vert \vec{p}_{a2} \vert},~~~~~P_{2}=\frac{\vec{p}_{b1}\cdot \vec{p}_{b2}}{ \vert \vec{p}_{b1} \vert\cdot\vert \vec{p}_{b2} \vert}.
\label{eq.6}
\end{eqnarray}

\noindent When\( \vert P_{1}\vert\)=\( \vert P_{2}\vert\)=1 means the maximum interferences between the upper and lower interband transitions, respectively. And \( \vert P_{1}\vert\)=\( \vert P_{2}\vert\)=0 represents no interference. Therefore, \( \vert P_{1}\vert\), \( \vert P_{2}\vert\) can be used to represent the quantum interference intensity in this proposed QD photocell. Current through the cell is given by \(j = e\Gamma \rho_{cc}\), while the voltage between levels \(|c\rangle\) and \(|v\rangle\) is given by\cite{Zhao2019,ZHAO2020106329},

\begin{eqnarray}
eV\!=\! E_{c}-E_{v}+k_{B}T_{a} \ln(\frac{\rho_{cc}}{\rho_{vv}}).
\label{eq.7}
\end{eqnarray}

Assuming that solar radiation and the phonon thermal source resonantly drives \(|b\rangle\) \(\rightarrow\) \(|\alpha_{3}\rangle\) \(\rightarrow\) \(|\alpha_{1}\rangle\), \(|b\rangle\) \(\rightarrow\) \(|\alpha_{2}\rangle\) \(\rightarrow\) \(|\alpha_{1}\rangle\) and \(|\alpha_{1}\rangle\) \(\rightarrow\) \(|c\rangle\), \(|v\rangle\) \(\rightarrow\) \(|b\rangle\) transitions. Therefore, the photoelectric conversion efficiency can be calculated as the ratio between the output power \(P_{out}\) and incident solar radiation power \(P_{s}\),

\begin{eqnarray}
\eta(\gamma_{a1,2}, \gamma_{b1,2}, n_{a1,2}, n_{1,2}) \!=\! \frac{P_{out}}{P_{s}}.
\label{eq.8}
\end{eqnarray}

\noindent where \(P_{out}\)=\(jV\), and  \(P_{s}\)=\(\frac{j}{e}\{\ln[\frac{(1+n_{a1})(1+n_{a2})(1+n_{b1})(1+n_{b2})}{n_{a1}n_{a2}n_{b1}n_{b2}}]\}\) with the fundamental charge of an electron being \(-e\).

\section{Results and analysis}

Here, the proposed QD photocell with two intermediate bands may be achieved by the confined electronic levels in self-assembled quantum dots\cite{Popescu2008Theoretical} with energy gap \(E_{c}-E_{v}\) being selected as 1.43 \(eV\) in the bulk GaAs band-gap. The parameters in this model may provide some experimental strategy for the multi-band QD solar cells. Owing to the quantum interference related to the intermediate transition rates, we will clarify the dependence of photoelectric conversion efficiency on the intermediate transition rates through different charge transfer channels. Next, we solve the diagonal and off-diagonal matrix elements via the steady-state solutions to Eqs.(\ref{eq.3}) analytically. Considering the cumbersome expression for the photo-to-charge efficiency \(\eta\), we follow the numerical quantitative approach to analyze the influence of quantum interference related to the intermediate transition rates on the photoelectric conversion.

\begin{figure}[!t]
\centerline{\includegraphics[width=0.3\columnwidth]{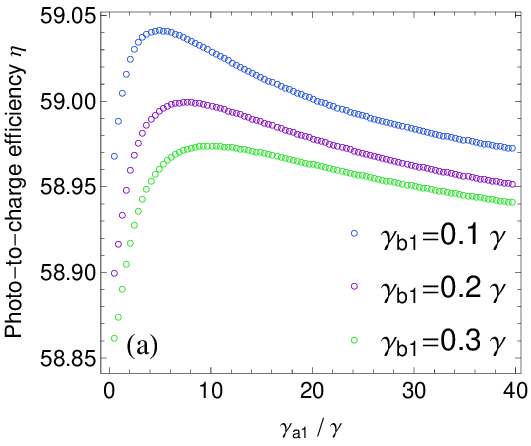}~\includegraphics[width=0.3\columnwidth]{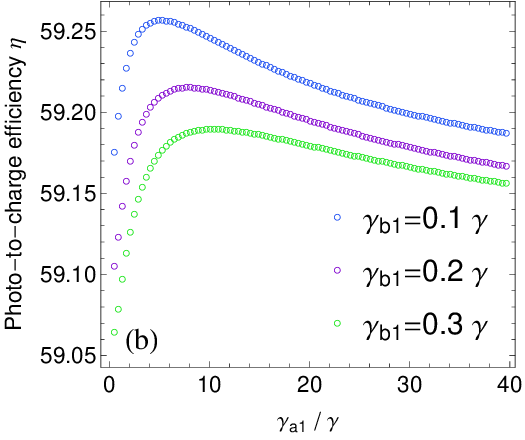}~\includegraphics[width=0.3\columnwidth]{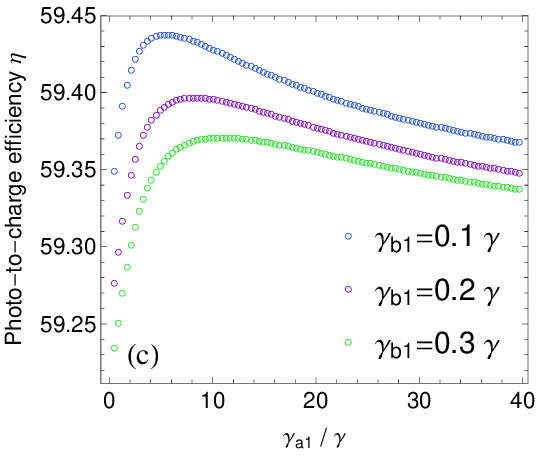}}
\center\caption{(Color online) Photoelectric conversion efficiency \(\eta\) optimized by the transition rates (\(\gamma_{a1}\), \(\gamma_{b1}\)) through the charge transfer channel \(|\alpha_{1}\rangle\)\(\leftrightarrow\) \(|\alpha_{2}\rangle\) \(\leftrightarrow\) \(|b\rangle\) with different solar photon numbers:(a) \(n_{a2}\)=4.5, (b) \(n_{a2}\)=5.0, (c) \(n_{a2}\)=5.5. Other parameters are \(n_{b1}\)=3\(n_{a1}\)=0.3, \(n_{b2}\)=0.50, \(n_{1}\)=3.5, \(n_{2}\)=4.5. \(\gamma_{a2}\)=0.25\(\gamma\), \(\gamma_{b2}\)=0.02\(\gamma\), \(\gamma_{1}\)=2.5\(\gamma\), \(\gamma_{2}\)=5.5\(\gamma\), \(\Gamma\)=0.45\(\gamma\), \(T_{s}=0.5ev\), \(T_{a}=0.0259ev\), \(p_{1}=0.45\), \(p_{2}=0.35\). And \(\gamma\) is the scale unit.}
\label{fig2}
\end{figure}

\begin{figure}[!t]
\centerline{\includegraphics[width=0.3\columnwidth]{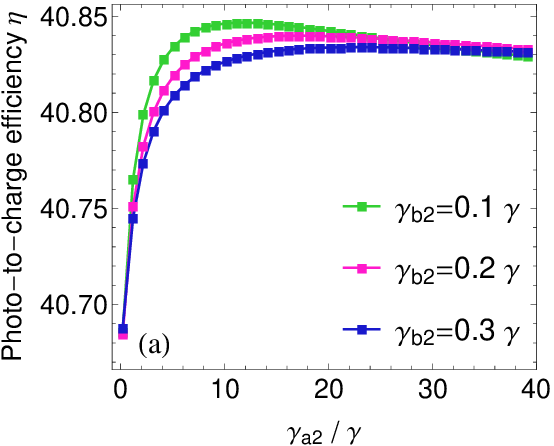}~\includegraphics[width=0.3\columnwidth]{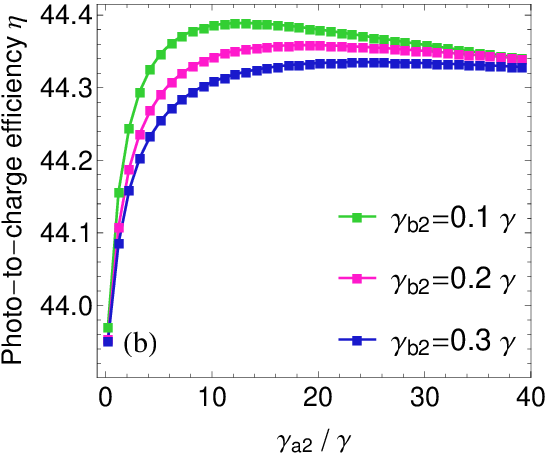}~\includegraphics[width=0.3\columnwidth]{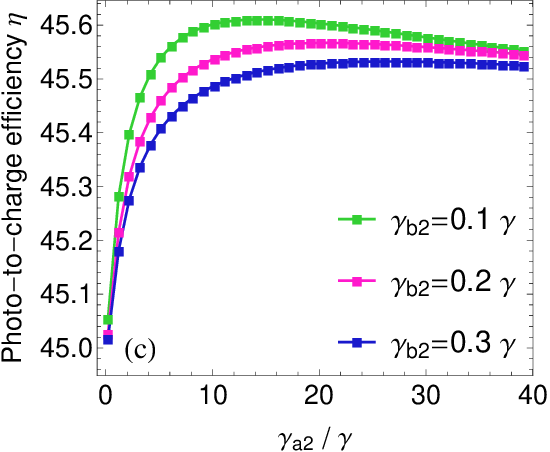}}
\center\caption{(Color online) Photoelectric conversion efficiency \(\eta\) optimized by the transition rates (\(\gamma_{a2}\), \(\gamma_{b2}\)) through the charge transfer channel \(|\alpha_{1}\rangle\) \(\leftrightarrow\) \(|\alpha_{3}\rangle\) \(\leftrightarrow\) \(|b\rangle\)  with different solar photon numbers: (a)\(n_{a1}\)=0.5, (b)\(n_{a1}\)=1.5, (c)\(n_{a1}\)=2.5 and \(n_{a2}\)=0.02, \(\gamma_{a1}\)=0.25\(\gamma\), \(\gamma_{b1}\)=0.02\(\gamma\). Other parameters are the same to those in Fig.\ref{fig2}.\label{fig3}}
\end{figure}

Absorbing more photons with energy below the bandgap is an efficient approach to enhance the photoelectric conversion. Hence, non-zero band splitting \(\Delta \) with different interband transition rates in different charge transfer channels attract our research interesting in this proposed QD photocell model. In the following, the efficiency \(\eta\) dependent transition rates with two non-degenerate intermediate bands \(|\alpha_{2}\rangle\), \(|\alpha_{3}\rangle\) (i. e., \(\Delta=0.1eV\)) are plotted in Fig.\ref{fig2}  and Fig.\ref{fig3} . Owing to the quantum interferences generated from multi-channel charge transfers, their underlying physical mechanism will be essentially affected by the interband transition rates in a quantum system. Hence, two charge transfer channels, i.e., \(|\alpha_{1}\rangle\) \(\leftrightarrow\) \(|\alpha_{2}\rangle\) \(\leftrightarrow\) \(|b\rangle\) and \(|\alpha_{1}\rangle\) \(\leftrightarrow\) \(|\alpha_{3}\rangle\) \(\leftrightarrow\) \(|b\rangle\) due to two non-degenerate intermediate bands  \(|\alpha_{2}\rangle\) and  \(|\alpha_{3}\rangle\), may bring about some peculiar quantum behaviors of the conversion efficiency in this proposed QD photocell.

Firstly, Fig.\ref{fig2} shows \(\eta\) as a function of the transition rates (\(\gamma_{a1}\), \(\gamma_{b1}\)) in the charge transfer channel \(|b\rangle\) \(\leftrightarrow\) \(|\alpha_{2}\rangle\) \(\leftrightarrow\) \(|\alpha_{1}\rangle\) with different solar photon numbers \(n_{a2}\). The results shows different roles of the conduction band transition rate \(\gamma_{a1}\) and intermediate band transition rate \(\gamma_{b1}\) in the photoelectric conversion efficiency \(\eta\). The efficiency \(\eta\) increases first and then decreases with the increment of the \(\gamma_{a1}\) in the range [0, 40\(\gamma\)] in Fig.\ref{fig2} from (a) to (c), but goes down with the increment of lower intermediate transition rate \(\gamma_{b1}\) by \(0.1 \gamma\). Not only that, but the absorbed solar photons plays a positive role in the photoelectric conversion efficiency \(\eta\). The peak efficiency \(\eta\) is promoted by the absorbed solar photons \(n_{a2}\). When \(\gamma_{b1}\)=\(0.1\gamma\), the maximum peaks approximately increase to 59.04\(\%\), 59.26\(\%\) and 59.44\(\%\) with \(n_{a2}\)=4.5, 5.0 and 5.5 in Fig.\ref{fig2}(a), (b) and (c), respectively.

What's the underlying physical mechanisms of the conduction band transition rate \(\gamma_{a1}\) in photoelectric conversion efficiency \(\eta\)?  As shown in Eqs.(\ref{eq.6}), the conduction band transition rate \(\gamma_{a1}\) is one reason for the quantum coherence generated from the upper transition rates. The increasing \(\gamma_{a1}\) about in the range [0, 8\(\gamma\)] enhances the quantum interference, which inhibits the photo-generated carriers radiating downward then facilitates the output charges. However, the continuously increasing \(\gamma_{a1}\) destroys the quantum interference effect, then, more excited carriers radiate directly to the intermediate band \(|\alpha_{2}\rangle\) simultaneously, which brings out the intrinsic loss in conversion efficiency \(\eta\) in Fig.\ref{fig2}.

However, the larger \(\gamma_{b1}\) means the shorter lifetime of carriers populating in the intermediate band \(|\alpha_{2}\rangle\). Hence, the photo-generated carriers radiative more rapidly to the valence band \(|b\rangle\) with a larger \(\gamma_{b1}\), which brings out the low occupation factor in the intermediate band \(|\alpha_{2}\rangle\). Hereupon, some researchers\cite{Yoshida2012Photon} have previously proposed a photon ratchet scheme to delay the lifetime of occupation in the intermediate state. Therefore, introducing a photon ratchet may be a clue to control \(\gamma_{b1}\) in this QD photocell for efficient photoelectric conversion. The express in Eqs.(\ref{eq.4}) illustrates that the larger amounts of absorbed photons \(n_{a2}\) can be achieved by a narrower gap, which is confirmed by the enhanced efficiency \(\eta\) with the increasing \(n_{a2}\) in Fig.\ref{fig2}.

Secondly, photoelectric conversion efficiency \(\eta\) is a function of another two transition rates (\(\gamma_{a2}\), \(\gamma_{b2}\)) in the charge transfer channel \(|b\rangle\) \(\leftrightarrow\) \(|\alpha_{3}\rangle\) \(\leftrightarrow\) \(|\alpha_{1}\rangle\) plotted in Fig.\ref{fig3}. The curves show the features of increasing first then decreasing with \(\gamma_{a2}\) and of monotonically decreasing with \(\gamma_{b2}\), as is similar to the curves in Fig.\ref{fig2}. Although the lower intermediate transition rate \(\gamma_{b2}\) shows the passive effect on \(\eta\), the absorbed solar photons \(n_{a1}\) plays an active role in \(\eta\) owing to the similar underlying physical mechanism to those in Fig.\ref{fig2}. However, the striking difference in the curves is the double-site with almost the same \(\eta\) but different \(\gamma_{b2}\) shown in Fig.\ref{fig3}. The results demonstrate that \(\gamma_{b2}\) loses its influence on the photoelectric conversion efficiencies \(\eta\) at about two identical \(\gamma_{a2}\) value sites. Obviously, it is not thoroughly to explain the underlying physical mechanism simply from the magnitude of \(\gamma_{b2}\). This should resort to the quantum interference generated from the upper transitions \(|\alpha_{1}\rangle\) \(\leftrightarrow\) \(|\alpha_{2,3}\rangle\).

\begin{figure}[!t]
\centerline{\includegraphics[width=0.3\columnwidth]{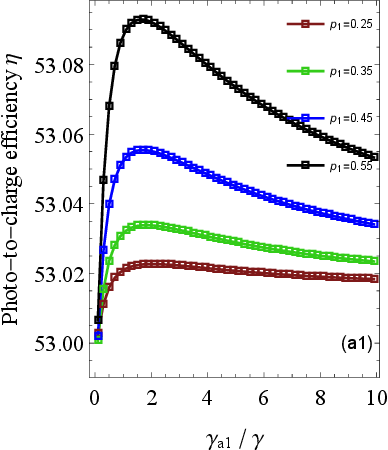}~\includegraphics[width=0.3\columnwidth]{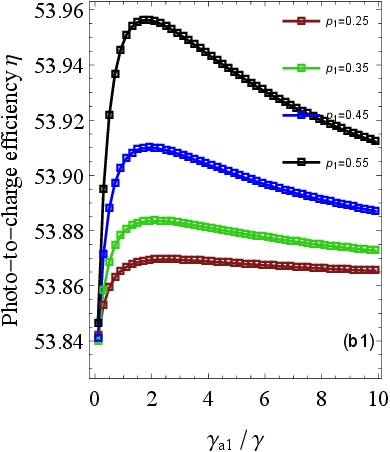}~\includegraphics[width=0.3\columnwidth]{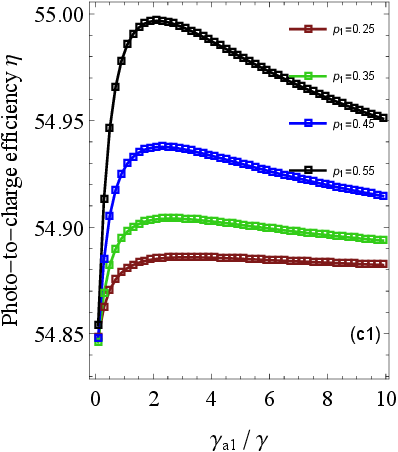}}
\centerline{\includegraphics[width=0.3\columnwidth]{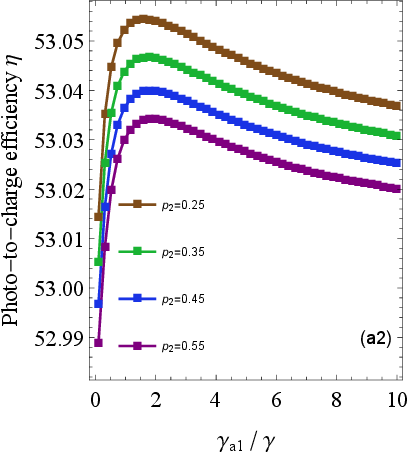}~\includegraphics[width=0.29\columnwidth]{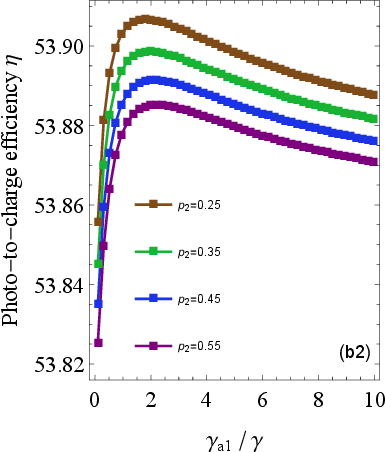}~\includegraphics[width=0.3\columnwidth]{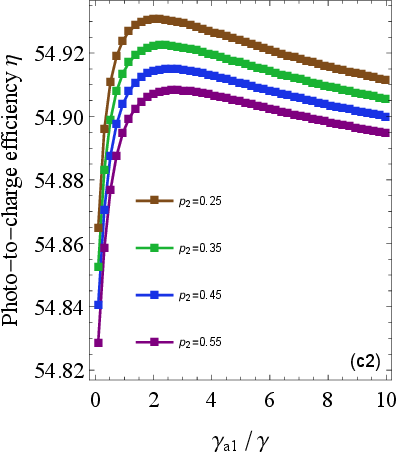}}
\caption{(Color online) Photoelectric conversion efficiency \(\eta\) optimized by the quantum interferences  \(P_{1}\), \(P_{2}\) generated from the upper and lower interband transitions, respectively. And the corresponding solar photon numbers are: Fig.a(1,2) \(n_{a2}\)=1.65, Fig.b(1,2) \(n_{a2}\)=2.15, Fig.c(1,2) \(n_{a2}\)=3.15. And \(n_{b1}\)=0.2, \(\Delta\)=0.1eV, the other parameters are the same as those in Fig.\ref{fig2}. }
\label{fig4}
\end{figure}

\begin{figure}[!t]
\centerline{\includegraphics[width=0.288\columnwidth]{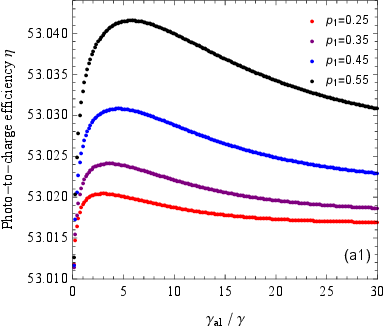}~\includegraphics[width=0.286\columnwidth]{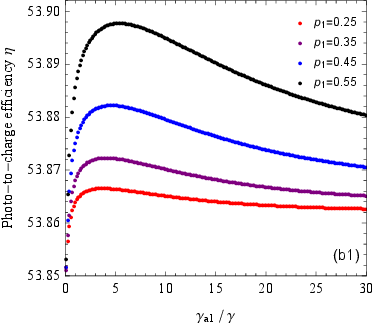}~\includegraphics[width=0.289\columnwidth]{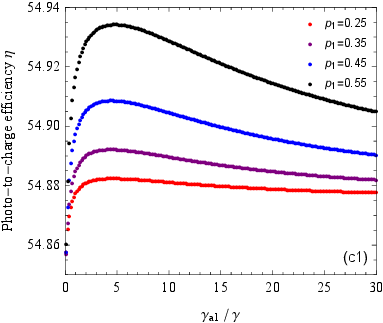}}
\centerline{\includegraphics[width=0.3\columnwidth]{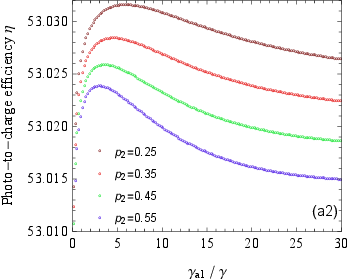}~\includegraphics[width=0.3\columnwidth]{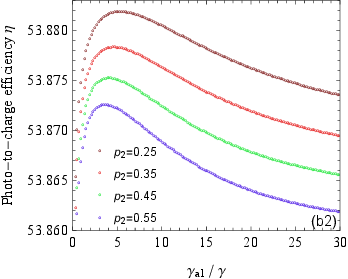}~\includegraphics[width=0.3\columnwidth]{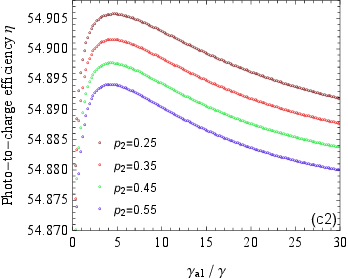}}
\caption{(Color online) Photoelectric conversion efficiency \(\eta\) optimized by the quantum interferences \(P_{1}\), \(P_{2}\) generated from the upper and lower intermediate band transitions, respectively. It takes the same parameters to those in Fig.\ref{fig4}, except \(\Delta\)=-0.5 \(eV\).}
\label{fig5}
\end{figure}

As a consequence, after the quantitative discussion between \(\eta\) and the transition rates (\(\gamma_{a1}\), \(\gamma_{b1}\)), (\(\gamma_{a2}\), \(\gamma_{b2}\)) in Fig.\ref{fig2} and Fig.\ref{fig3}, the quantum interferences generated from the upper transitions \(|\alpha_{1}\rangle\) \(\leftrightarrow\) \(|\alpha_{2,3}\rangle\) and lower transitions \(|\alpha_{2,3}\rangle\) \(\leftrightarrow\) \(|b\rangle\) are explored explicitly with two different band splitting \(\Delta\)=0.1eV and -0.5eV in Fig.\ref{fig4} and Fig.\ref{fig5}, respectively.
Fig.\ref{fig4} plots efficiency \(\eta\) optimized by the quantum interference intensity \(P_{1}\) with a constant \(P_{2}\)=0.4 in Fig.\ref{fig4}(a1), (b1) and (c1), while the efficiency \(\eta\) versus the quantum interference intensity \(P_{2}\) are shown in Fig.\ref{fig4}(a2), (b2) and (c2) with the constant \(P_{1}\)=0.4. Comparing the curves in Fig.\ref{fig4} ($\Phi$1) and ($\Phi$2)\(_{\Phi=a,b,c}\), two significant conclusions are drawn: one is that the differences of different peaks optimized by \(P_{1}\) are larger than those optimized by \(P_{2}\) under the same conditions, which highlights the regulatory sensitivity of \(P_{1}\) on the conversion efficiency. The other one is \(P_{1}\), which shows a positive role in the output efficiency \(\eta\): the output efficiency \(\eta\) gradually increases with \(P_{1}\) while a negative control function shown by \(P_{2}\), \(\eta\) decreases with \(P_{2}\).

The underlying mechanism, for the enhanced \(\eta\) in Fig.\ref{fig4}(a1), (b1) and (c1)) owing to quantum coherence from the upper transition \(|\alpha_{1}\rangle\) \(\leftrightarrow\) \(|\alpha_{2,3}\rangle\) is similar to the aforementioned works\cite{Scully2010Quantum,Svidzinsky2011Enhancing,Daryani2017High}, in which the output power can be enhanced by the quantum coherence scheme. However, the inhibited \(\eta\) by \(P_{2}\) in Fig.\ref{fig4}(a2), (b2) and (c2)) is due to the shorter lifetime of carriers in the intermediate bands \(|\alpha_{2,3}\rangle\), which exacerbates the radiation of photo-generated charges back to the ground state \(|b\rangle\). As is a different physical regime caused by the quantum coherence from the upper transition \(|\alpha_{1}\rangle\) \(\leftrightarrow\) \(|\alpha_{2,3}\rangle\).

In order to clarify quantum interference dependent the band splitting \(\Delta\), Fig.\ref{fig5} plots the efficiency \(\eta\) optimized by the quantum interference strength \(P_{1}\), \(P_{2}\) with \(\Delta\)=-0.5 \(eV\). As shown in Fig.\ref{fig5}, the curves display show small difference except the slightly smaller peaks comparing with those in Fig.\ref{fig4}. The maximum peak is about 54.935\(\%\) with \(n_{a2}\)=3.15 in Fig.\ref{fig5}(c1) and the minimum peak is 53.024\(\%\) with \(n_{a2}\)=1.65 Fig.\ref{fig5}(a2), respectively. The peak values are smaller than those in Fig.\ref{fig4} under the same physical condition except the band splitting \(\Delta\), it concludes that this QD photocell with small band splitting can produce slightly larger photoelectric conversion efficiency than that with a larger \(\Delta\).

According to the above conclusions, the key issue in this proposed scheme is the flexible control of intermediate band transition rates. In the photon rachet scheme\cite{Yoshida2012Photon}, a controllable lifetime of the intermediate band was provided by the incorporated photon rachet band, which may be an inspiration to manipulate the transition rates for this proposed QD photocell scheme. However, in order to highlight the control of intermediate band transition rates on photoelectric conversion efficiency, this work does not consider other influences on photoelectric conversion efficiency, such as the carriers transport mechanism and influence mechanism of the ambient environment, and other internal or external aspects. But, we argue that these mentioned points do not influence to reveal the significance of the current conclusions. And we believe the future work will bring some important insights into the above mentioned issues.

\section{Conclusion}

In conclusion, this work discussed the dependence of photoelectric conversion on the quantum coherence related different transition rates, and proposed some optimizations for photoelectric conversion efficiency via the intermediate band transition rates in a QD photocell with two intermediated bands.  The characteristic of first increasing and then decreasing generated by \(\gamma_{a1}\), \(\gamma_{a2}\) in conversion efficiency is achieved by the upper quantum coherence strength \(P_{1}\) in both two different charge transfer channels \(|b\rangle\) \(\leftrightarrow\) \(|\alpha_{2}\rangle\) \(\leftrightarrow\) \(|\alpha_{1}\rangle\) and \(|b\rangle\) \(\leftrightarrow\) \(|\alpha_{3}\rangle\) \(\leftrightarrow\) \(|\alpha_{1}\rangle\). Their underlying physical mechanisms come from the quantum coherence generated by the upper transition rates (\(\gamma_{a1}\) , \(\gamma_{a2}\)), which constructs, subsequently destructs the quantum coherence during the increasing processes of (\(\gamma_{a1}\), \(\gamma_{a2}\)). However, the increments of two lower intermediate band transition rates \(\gamma_{b1}\), \(\gamma_{b2}\) inhibit the conversion efficiency in both of above two different charge transfer channels. These results are proved by the quantum coherence strength \(P_{2}\) generated from the two lower intermediate band transitions, and by the decreasing peak values of efficiencies \(\eta\). Different roles of quantum interferences induced by the upper and lower intermediate band transitions, may point out some new experimental directions to enhance the photoelectric conversion in the QD photocell.

\begin{acknowledgments}
We thank the financial supports from the National Natural Science Foundation of China ( Grant Nos. 62065009 and 61565008), and Yunnan Fundamental Research Projects, China ( Grant No. 2016FB009 ).
\end{acknowledgments}






\bibliography{references}
\bibliographystyle{unsrt}
\end{document}